\documentclass[aps,prx,twocolumn,showpacs,superscriptaddress,groupedaddress]{revtex4}

\usepackage{graphicx}
\usepackage{dcolumn}
\usepackage{bm}
\usepackage{amssymb}
\usepackage{soul}
\usepackage{color}

\def\be{\begin{equation}}
\def\ee{\end{equation}}
\def\bea{\begin{eqnarray}}
\def\eea{\end{eqnarray}}

\begin{document}

\title{Proximate transition temperatures amplify linear magnetoelectric coupling in strain-disordered multiferroic BiMnO$_{3}$}
\author{Patrick~R.~Mickel}\affiliation{University of Florida, Gainesville, Florida, 32611, USA}\affiliation{Sandia National Laboratories,P.O. Box 5800, Albuquerque, New Mexico 87185,USA}
\author{Hyoungjeen Jeen}\affiliation{University of Florida, Gainesville, Florida, 32611, USA}\affiliation{Pusan National University, Pusan 609735, South Korea}
\author{Pradeep Kumar}\affiliation{University of Florida, Gainesville, Florida, 32611, USA}
\author{Amlan Biswas}\affiliation{University of Florida, Gainesville, Florida, 32611, USA}
\author{Arthur~F.~Hebard} \email[Corresponding author:~]{afh@phys.ufl.edu} \affiliation{University of Florida, Gainesville, Florida, 32611, USA}

\begin{abstract}

We report a giant linear magnetoelectric coupling in strained BiMnO$_{3}$ thin films in which the disorder associated with an islanded morphology gives rise to extrinsic relaxor ferroelectricity that is not present in bulk centrosymmetric ferromagnetic crystalline BiMnO$_{3}$. Strain associated with the disorder is treated as a local variable which couples to the two ferroic order parameters, magnetization $\vec{M}$ and polarization $\vec{P}$.
A straightforward \lq \lq gas under a piston\rq \rq \thinspace thermodynamic treatment explains the observed correlated temperature dependencies of the product of susceptibilities and the magnetoelectric coefficient together with the enhancement of the coupling by the proximity of the ferroic transition temperatures close to the relaxor freezing temperature. Our interpretation is based on a trilinear coupling term in the free energy of the form $\vec{L} \cdot (\vec{P} \times \vec{M})$ where $\vec{L}$ is a hidden antiferromagnetic order parameter, previously postulated by theory for BiMnO$_{3}$. This phenomenological invariant not only preserves inversion and time reversal symmetry of the strain-induced interactions but also explains the pronounced linear magnetoelectric coupling without using the more conventional higher order biquadratic interaction proportional to $(\vec{P} \cdot \vec{M})^2$.  
\end{abstract}

\maketitle
\date{\today}

\section{INTRODUCTION}

Magnetoelectric (ME) coupling, the induction of electric (magnetic) polarization via magnetic (electric) fields, represents an area of intense research due to both its complex physical origins and its potential technological applications \cite{Revival,ME1,ME3}. According to general thermodynamic considerations, multiferroics - materials which simultaneously possess spontaneous magnetic and electric polarizations - have long been predicted as prime candidates for enhanced ME coupling\cite{ME60s,ME2,dzya}. Additionally, strong limitations on ME coupling, such as mutually exclusive requirements in electronic configurations for each ferroic order, have been identified\cite{d0}. Accordingly, understanding global energetic constraints on ME coupling - and ways to relax these constraints in multiferroic systems - is of fundamental importance for this class of materials to reach their potential. In particular, the use of concomitant ferroic transition temperatures to amplify ME coupling has been proposed\cite{dzya,boyd}, but an experimental study providing insight into the thermodynamic constraints on such an enhancement has yet to be presented.

Crystalline BiMnO$_{3}$ (BMO) is recognized as the only insulating perovskite that is strongly ferromagnetic. Early indications that BMO should also be a displacive ferroelectric earned it the distinction of being called the ``hydrogen atom" of multiferroics\cite{HillRabe}. Orbital ordering of Mn spins was determined to be responsible for ferromagnetism\cite{BMO_OO}, and a structural distortion induced by the lone pair of the Bi ions was believed to produce spontaneous electric polarization\cite{Vis6s}. However, subsequent neutron scattering and density functional theory (DFT) investigations have contested the non-centrosymmetric classification of BiMnO$_3$, casting doubt over its intrinsic ferroelectric properties \cite{Baettig,Belik,Scott2}. 

Despite these claims of a centrosymmetric crystal structure for bulk BMO, reports of ferroelectricity in thin films of BiMnO$_3$ remain\cite{DosSantos,Sharan,Son,Gajek,STObuf,Jeen,DeLuca}, suggesting alternate extrinsic mechanisms are present. In a comprehensive review of polar and nonpolar phases of BiMnO$_3$ Belik\cite{BMOreview} has pointed out that any particular BMO thin-film sample can have strain-sensitive structural and compositional modifications that are strongly correlated with the appearance of ferroelectricity. Importantly, the film resistivity must be sufficiently high to insure that leakage currents do not reduce the charge separation associated with remnant electric dipole moments. 

We show that in our pulsed laser deposited multiferroic BiMnO$_3$ thin films the operating \textit{extrinsic} mechanism that gives rise to a pronounced and surprisingly large linear magnetoelectric effect is strain disorder. We observe temperature-dependent transitions of the magnetic ($\vec{M}$) and ferroelectric ($\vec{P}$) order parameters which are in close proximity to each other and give rise to a strongly enhanced magnetoelectric coupling in the vicinity of the transition temperatures. The interacting ferroic order parameters and their contribution to a large magnetoelectric effect are described using a strain-dependent trilinear interaction term in the free energy of the form $\vec{L} \cdot (\vec{P} \times \vec{M})$. To preserve inversion and time reversal symmetry, a staggered order parameter $\vec{L}$ representing an antiferromagnetic background is used. The trilinear form of the interaction originally discussed by Fennie \cite{Fennie} is motivated by a Dzyaloshinsky-Moriya\cite{Dzy,Moriya} type antisymmetric interaction.  Recent theoretical treatments suggest that \lq \lq hidden\rq \rq antiferromagnetism must be present in BiMnO$_3$\cite{hiddenAFM,supertetragonal}. Using the more conventional higher-order biquadratic interaction, $(\vec{P} \cdot \vec{M})^2$, which is also symmetry allowed, does not explain the data as will be explained below.

\section{Methods}

Our multiferroic samples are 60 nm-thick thin films of stoichiometric BiMnO$_{3}$ grown using pulsed laser deposition on (001) SrTiO$_{3}$
(STO) substrates. X-ray characterization showed strong peaks demonstrating coherent local order (see Ref. \cite{Jeen} for stoichiometric analysis and growth parameters). The films grow with a 001 orientation (pseudo-cubic notation) on STO substrates with a compressive strain due to lattice mismatch of 0.77\%. Optimized tuning of deposition rate, substrate temperature, target stoichiometry, oxygen partial pressure and post deposition cooling rates leads to films with no discernible impurity phases and resistivities higher than 10$^6$ $\Omega$\thinspace cm\ in the temperature range where all measurements were taken \cite{Jeen}. 

\begin{figure}
\includegraphics[angle=0,width=0.5\textwidth]{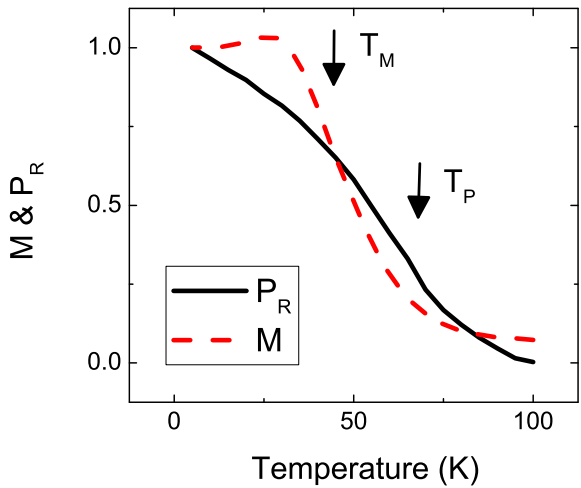}
\caption{Temperature dependence of remanant polarization $P_R$ (solid black) and magnetization $M$ (dashed red) demonstrating the near overlap of the ferroelectric and magnetic transitions. The vertical arrows show the temperatures where the maxima in the slopes occur.}
\label{figure1}
\end{figure}

The surface topography of the films imaged using tapping-mode atomic force microscopy (AFM) shows a three-dimensional island growth mode with an rms roughness of approximately 10~nm. Accordingly, nonuniform strain with high values of strain at the island edges is expected\cite{Chen,Biswas}. This scenario is consistent with temperature-dependent magnetization measurements acquired with a Quantum Design SQUID magnetometer using an in-plane field of 500~Oe\cite{Jeen}. The magnetization shown in Fig.~\ref{figure1} (dashed red line) shows an onset near 85~K and a saturated moment of approximately 1~$\mu_B$/Mn. The transition and saturated moment in our thin films is lower than the transition ($T_c \approx 105$~K) and saturated moment (3.6 $\mu_B$/Mn) obtained in polycrystalline samples\cite{DosSantos,Chiba}. Although the presence of Bi vacancies may contribute to reduced magnetism in BMO\cite{Gajek}, we attribute our reductions of transition temperature and saturated moment primarily to the substrate induced strain and the strain disorder associated with the islanded morphology.

The remanent ferroelectric polarization was measured using a Radiant Technologies ferroelectric tester and a previously-described interdigital capacitor geometry\cite{Jeen}. The polarization in the remanent hysteresis loop is calculated by isolating the transferred charge due solely to domain-switching. 

\begin{figure}
\includegraphics[angle=0,width=0.55\textwidth]{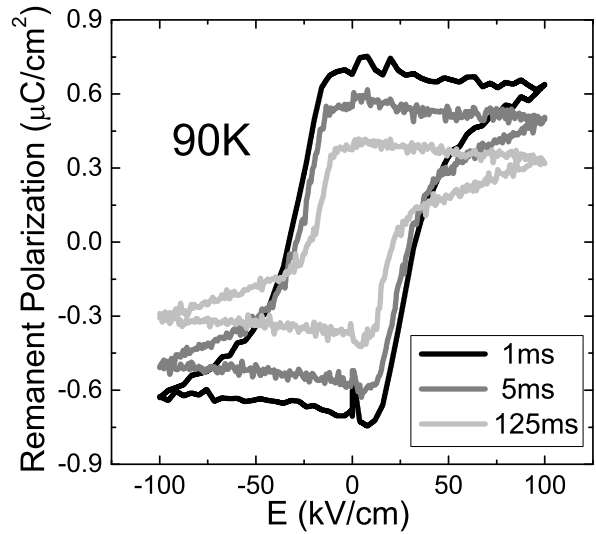}
\caption{Polarization loops at 90~K acquired on the time scales indicated in the legend show relaxor ferroelectric behavior. On the longer time scales the remnant polarization has more time to interact with the thermal background and decrease in magnitude.}
\label{relaxorloops}
\end{figure}

The ferroelectric polarization loops shown in Fig.~\ref{relaxorloops} shrink in area as the time (indicated in the legend) to complete a loop increases. This behavior is a signature of relaxor ferroelectricity in which the charge separation associated with induced polarizations decays with time because of interactions with the phonons of the thermal background. The remanant polarization $P_R$ for each loop is taken as the polarization at an applied electric field $E = 0$ where the horizontal portion of each loop crosses the vertical axis. Dielectric characterizations of these same films\cite{Mickel1} show two thermally-activated loss peaks that are signatures of relaxor ferroelectricity with the real part of the dielectric constant becoming frequency independent and diverging at a relaxor freezing temperature\cite{Vogel} $T_F$ near 70~K. For $T > T_F$ any remanence in the sample will decay to zero on laboratory time scales.

Since Bi-doped SrTiO$_{3}$ has been reported to be a relaxor ferroelectric\cite{Ang}, we necessarily should be concerned with possible ferroelectric contributions from Bi doping of the SrTiO$_{3}$ substrate during the high temperature growth (632$^{\circ}$C) \cite{Jeen} of the BiMnO$_3$ films. This source of ferroelectricity was ruled out via two methods. First a BiMnO$_{3}$ film was grown and then etched with a KI solution. After depositing interdigital electrodes directly on the exposed substrate, remanent polarization measurements revealed distorted hysteresis loops which were more than 100$\times$ smaller than those measured for BiMnO$_{3}$.  In addition, dielectric measurements did not display the relaxations observed in the BiMnO$_{3}$ films. In our second method, BiMnO$_{3}$ films were grown on alternate substrates (NdGaO$_{3}$ and SrLaGaO$_{4}$) which have different composition from that of SrTiO$_{3}$. Our measurements showed reduced but open remanent polarization loops as expected since the mismatch strain is less. Accordingly, Bi-doped SrTiO$_{3}$ was definitely ruled out as the source of relaxor ferroelectricity reported here.

\section{Results and Discussion}
\label{sec:results and discussion}

The proximity of the magnetic and ferroelectric transitions as measured by the overlap of the rapid changes in the temperature-dependent magnetization $M(T)$ (the order parameter for the magnetic transition) and the remanant polarization $P_R(T)$ is shown in Fig.~\ref{figure1}. The relaxor (diffuse) nature of the ferroelectric transition shown in Fig.~\ref{relaxorloops} implies the presence of disorder-induced precursor polar microregions which with decreasing temperature grow in size and fluctuate on longer time scales\cite{Tsurumi}. In the temperature regions of Fig.~\ref{figure1} where there is a pronounced increase in the magnitude of $P_R$ with decreasing temperature, the polarizations become increasingly stable in time. The question of whether the proximity of the two transitions is accidental will be addressed below. 

\begin{figure}
\includegraphics[angle=0,width=0.55\textwidth]{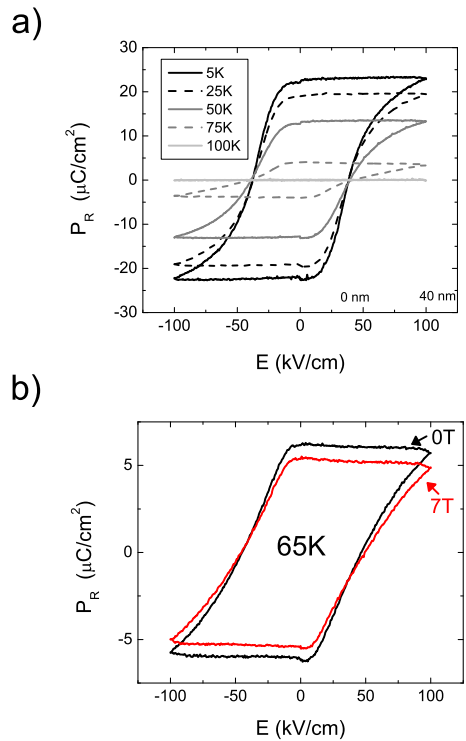}
\caption{Temperature and magnetic field dependence of electric polarization loops. \textbf{(a)} Hysteretic polarization loops at the indicated temperatures span a temperature range of $\approx$ 100\thinspace K with a maximum remanent polarization, $P_{R}$, of $\approx$\thinspace 23\thinspace $\mu$\thinspace C/cm$^{2}$ at 5\thinspace K. \textbf{(b)} A magnetic field of 7\thinspace T (red) is shown to decrease the FE polarization by $\approx$ 10\% (red) from its value in zero field (black).} 
\label{figure3}.
\end{figure}

As shown in Fig.~\ref{figure3}a the polarization loops measured at the temperatures indicated in the legend steadily increase in area as the temperature is reduced. The remanent-polarization ($P_{R}$) evaluated at the electric field $E = 0$ appears near 110$\thinspace$K and increases slowly to a large value of 23\thinspace$\mu$C/cm$^{2}$ at 5\thinspace K. Application of an external tensile strain on the order of 0.01\% using a three-point \lq \lq beam-bending\rq \rq technique increases $P_R$ at 5~K by 50\%\cite{Mickel1}. Our experimental observations of relaxor ferroelectricity, which requires the presence of disorder-induced polar micro regions\cite{Tsurumi}, together with the above mentioned sensitivity of $P_R$ to externally applied strain strongly suggests that disordered strain regions, which are not present in crystalline bulk BMO, are responsible for the robust ferroelectricity observed in thin-film BMO.    

\begin{figure}
\includegraphics[angle=0,width=0.5\textwidth]{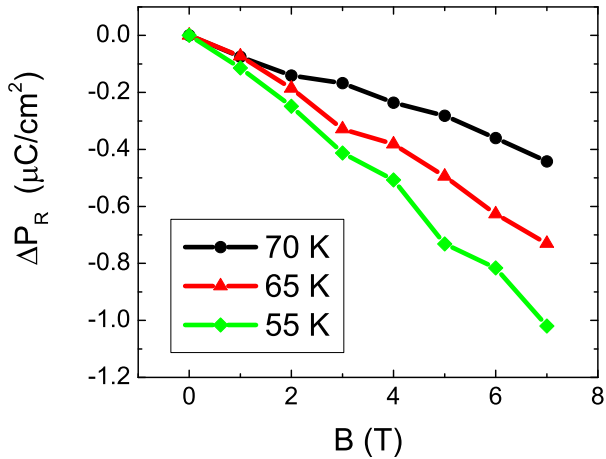}
\caption{The remnant polarization $P_R$ decreases at all temperatures with a linear dependence on magnetic field. The magnetoelectric coupling constants $\tilde{\alpha}(T)$ for the three selected temperatures in the legend are extracted from the slopes of each of these curves. The maximum $\tilde{\alpha}(T)$with a value of -0.1 $\mu$C/cm$^2$T (-1.25 ns/m) occurs near 55~K.}
\label{figure4}
\end{figure}

In addition to the robust extrinsic ferroelectricity, we also observe a strong magnetoelectric (ME) coupling which is especially pronounced in the region where the transitions in $P_R$ and $M$ overlap. The lower panel of Fig.~\ref{figure3} shows how the application of a 7~T magnetic field shrinks the area of the polarization loop and reduces $P_R$ by $\approx$10\%. The reduction of $P_R$ is linearly dependent on the applied magnetic field $B = \mu_0 H$ as shown in the selected curves of Fig.~\ref{figure4} for the temperatures indicated in the legend. We find that the ME effect, characterized by a coupling coefficient defined by the slope $\tilde{\alpha} = - \partial P_R/ \partial (\mu_0 H)$, is negative and linear at all temperatures. The coupling coefficient is found to be independent of field direction and is quite large, reaching a value of -0.1~$\mu$C/cm$^{2}$T (-1.25 ns/m in SI units) at $T = 65$~K.  This value is approximately 35 times larger than the current record for single-phase linear ME coupling of 36.7 ps/m found in TbPO$_{4}$\cite{Revival,TbPO4}. Coupling larger than that presented here has been reported, but strong non-linearities have made such couplings difficult to interpret thermodynamically\cite{Spirals,Scott}. In addition, $\tilde{\alpha}$ averages out the effect of disorder in our sample, since the measurement length scales for both $M$ and $P$ are much larger (an order of magnitude or more) than the typical island size and separation\cite{Jeen}. The relevance of this point will be discussed in section C.

\subsection{Gas under a piston thermodynamics}

We have chosen the relevant thermodynamic variables describing our multiferroic system to be the order parameters $M$ and $P$ rather than the magnetic induction $B$ and the electric displacement $D$, since the 
\lq \lq medium in an external field\rq \rq \thinspace is analogous to the \lq \lq  gas under a piston\rq \rq \thinspace system\cite{Mikaelyan}. By making the analogy to a perfect gas described by the state variables \textit{p} (pressure), \textit{V} (volume) and \textit{T} (temperature), we incorporate the intensive variables, \textit{M} and \textit{P}, into the free energy by replacing the work term $pdV$ with the substitutions\cite{Rushbrooke,Mikaelyan} $p \rightarrow H$, $V \rightarrow -\mu_0 MV = -\mu_0 \chi_M HV$ for the magnetic energy and $p \rightarrow E$, $V \rightarrow -PV = -\epsilon_0 \chi_P E V$ for the electric energy. SI units are used throughout.
 
These substitutions become particularly useful in finding expressions for the electric and magnetic susceptibilities beginning with the well-known expression for the heat capacity difference $C_p - C_V$ of a $pVT$ system ,
\begin {equation}
C_p - C_V = -T(\partial V/\partial T \vert_p)^{2}/(\partial V/\partial p)_T ~~~,
\label{thermo}
\end{equation}  
which reduces to $C_p - C_V = R$ for one mole of a perfect gas with $R$ equal to the gas constant.
By using the constitutive relations, $\chi_M = (\partial M/ \partial H)_T$, and $\chi_P = \epsilon_{0}^{-1}(\partial P/ \partial E)_T$, to define the dimensionless susceptibilities $\chi_M$ and $\chi_P$ for the order parameters \textit{M} and \textit{P} respectively, we incorporate these variable replacements and accompanying constitutive relations into Eq.~\ref{thermo} and arrive at the thermodynamic relations:
\bea
\chi_M(T) &=& \mu_0 T(\partial M(T)/\partial T \vert_H)^{2}/(c_H - c_M) \cr 
\cr
\chi_P(T) &=& \epsilon_0^{-1} T(\partial P(T)/\partial T \vert_E)^{2}/ (c_E - c_P) ~.
\label{susceptibilities}
\eea%
Since the isochoric specific heats, $c_{H,M} = C_{H,M}/V$ and $c_{E,P} = C_{E,P}/V$, have a very weak temperature dependence in the vicinity of the respective phase transition temperatures $T_M$ and $T_P$\cite{Binney}, the above expressions of Eq.~\ref{susceptibilities} reduce to the simple proportionalities:   
\bea
\chi_M(T) &\propto& T_M (\partial M(T)/\partial T \vert_H)^{2} \cr
\cr
\chi_P(T) &\propto& T_P(\partial P(T)/\partial T \vert_E)^{2} ~.
\label{proportionalities}
\eea
We emphasize that these expressions for $\chi_M (T)$ and $\chi_P (T)$ are derived from thermodynamics and do not require $M$ and $P$ to be linearly dependent on $H$ and $E$ respectively. 

\subsection{Angle dependence of the Helmholtz free energy}

Using these same variable substitutions we write the total free energy as a sum of terms, 
\be
F(P,M) = F_P (P) + F_M (M) +  F_{int}
\label{free_energy}
\ee
where $F_P(P)$ and $F_M(M)$ include quadratic terms in $\vec{P}$ and $\vec{M}$ respectively, i.e.,  
\be
F_P (P) + F_M (M) = \left( \frac{1}{2 \epsilon_0 \chi_P}\right) P^2 + 
\left( \frac{\mu_0}{2\chi_M} \right) M^2%
\label{Helmholtz1}  
\ee
and $F_{int}$ is an interaction term including the symmetry-allowed trilinear and biquadratic terms, i.e.,  
\be
F_{int} =  a_1 \vec{L} \cdot (\vec{P} \times \vec{M}) + a_2 (\vec{P} \cdot \vec{M})^2 ~ ~.
\label{Helmholtz2}
\ee
where $a_1$ and $a_2$ are positive constants. In Eq.~\ref{Helmholtz1} we have omitted higher order quartic terms in $P^4$ and $M^4$ since, as will be shown below, linear response provides a good description of our data.

We next minimize the angular part of the free energy 
\be
F_{int} =  a_1 L P M \sin \theta + a_2 (PM)^2 \cos^2 \theta ~ ~.
\label{Fint_theta}
\ee
where $\theta$ is the angle subtended by $\vec{P}$ and $\vec{M}$, $L$ is redefined as the projection of $\vec{L}$ on $\vec{P} \times \vec{M}$, and $P = |\vec{P}|$ and $M = |\vec{M}|$ will henceforth be used as parametric scalar variables having their own temperature and field dependencies. The extrema of Eq.~\ref{Fint_theta} at $\theta = \theta_0$ are found by setting $\partial F_{int}/\partial \theta = 0$, leading to two sets of solutions for $\theta_0$; namely that coming from $\cos \theta_0 = 0$, i.e., $\theta_0 = \pi/2, 3 \pi/2$, or $\sin \theta_0 = a_1 L/2 a_2 PM < 1$. The free energy is found to be a minimum (i.e., $\partial^2F_{int}/ \partial \theta^2 > 0)$ only for the specific cases $\theta_0 = \pi/2$, provided $a_1L/2 a_2 PM < 1$, and $\theta_0 = 3\pi/2$, with no restrictions on the ratio $a_1L/2 a_2 PM$. For the solution $\theta_0 = \sin^{-1}(a_1 L/2 a_2 PM)$ with $a_1 L/2 a_2 PM < 1$ the free energy is a maximum. Accordingly, the only two possibilities giving a minimum in the free energy are $\theta_0 = \pi/2, 3 \pi/2$. Substitution of these angles into Eq.~\ref{Fint_theta} reveals that $\theta_0 = 3 \pi / 2$ gives a global minimum,
$F_{int} = - a_1 LPM$, so that the total free energy of Eqs. ~\ref{free_energy} and~\ref{Helmholtz2} at $\theta_0 = 3 \pi/2$ simplifies to 
\be
F(P,M) = \left( \frac{1}{2 \epsilon_0 \chi_P} \right) P^2 + \left( \frac{\mu_0}{2\chi_M} \right) M^2 -\mu_0\alpha PM 
\label{Helmholtz3}  ,
\ee
where the positive constant $\alpha = a_1 L/\mu_0$ is proportional to $L$ and therefore breaks both time and inversion symmetry. This form for the minimum free energy holds with no restrictions on the ratio $a_1L/2 a_2 PM$, thereby allowing the values of $P$ and $M$ to be small.

\subsection{Strain and a hidden antiferromagnetic order parameter?}
\label{subsec:a hidden antiferromagnetic order parameter}

A previous treatment of BMO using Landau theory incorporates quartic, $P^4$ and $M^4$, and biquadratic, $(PM)^2$, terms in the free energy\cite{Howczak}. In contrast, our finding in the above analysis that the angle-averaged free energy is minimum at $\theta _0 = 3\pi /2$ implies that the conventional biquadratic coupling term ($\vec{P} \cdot \vec{M})^2$ in Eq.~\ref{Helmholtz2} is zero and hence can be ignored when the trilinear term  $\vec{L} \cdot (\vec{P} \times \vec{M})$ is present. Having taken account of the angular dependence the resulting free energy expressed in Eq.~\ref{Helmholtz3} is considerably simplified with a bilinear interaction proportional to $PM$ and at the same time symmetry preserving because of the absorption of $L = |\vec{L}|$ into the constant $\alpha$. In the following analysis we therefore consider $P$ and $M$ to be the principal order parameters with $L$ playing an important but sub-dominant (\lq \lq hidden\rq \rq \cite{hiddenAFM}) role.

To explicitly include strain, we recall that pristine crystals of BiMnO$_3$ are centrosymmetric \cite{Baettig,Belik,Scott2} and thus do not undergo ferroelectric transitions. In contrast our films exhibit a \lq \lq strain-disordered\rq \rq ~islanded morphology, confirmed by AFM scans \cite{Jeen,Mickel1}, together with a pronounced ferroelectric transition (see Fig.~\ref{figure3}). 
Accordingly, we conclude from these observations that strain described by a local strain variable $\delta \acute{x}$ associated with disorder, which is present only in films but not in crystals, is the source of an extrinsic mechanism for ferroelectricity in BiMnO$_3$ thin films. 

This line of reasoning is pursued by making the ansatz that local distortions described by $\delta \acute{x_i}$ couple linearly to site specific moments, $p_i$ and $m_i$, to give the interaction term of Eq.~\ref{Helmholtz3}. More specifically, we now write the interaction contribution to the free energy in terms of disorder-related strain terms,
\be
\delta F_{int} =  \acute{g_1} p_i \delta \acute{x_i} + \acute{g_2} m_i \delta \acute{x_i} + \acute{k}\delta \acute{x_i}^2/2  ,
\label{strainFE}
\ee
which when evaluated at equilibrium, i.e., $\delta \acute{x_i} = -(\acute{g_1} p_i + \acute{g_2} m_i)/\acute{k}$, are found to lower the free energy by an amount
\be
F_{int} = -\acute{k}\delta \acute{x_i}^2/2 = -(\acute{g_1} p_i +\acute{g_2} m_i)^2/2\acute{k}  .
\label{strainFE_v2}
\ee
For clarity, the coupling constants $\acute{g_1}$ and $\acute{g_2}$ and the restoring force constant $\acute{k}$ associated with strain disorder are primed. Expansion of the quadratic form in Eq.~\ref{strainFE_v2} and averaging over all sites gives rise to corrections $-\acute{g_1}^2/4\acute{k}$ and $-\acute{g_2}^2/4\acute{k}$ to $P^2$ and $M^2$ respectively in Eq.~\ref{Helmholtz1}, which can be ignored, and to the additional bilinear interaction term $\acute{g_1} \acute{g_2}PM/\acute{k}$. Using this model we can therefore make the identification $\mu_0\alpha = a_1L = \acute{g_1} \acute{g_2}/\acute{k}$ in Eq.~\ref{Helmholtz3}. 

The strain induced corrections to the free energy must preserve both inversion and time reversal symmetry. This is true for the 
$\acute{g_1} p_i \delta \acute{x_i}$ and $\acute{k}\delta \acute{x_i}^2/2$ terms of our above ansatz but not for the magnetic coupling term, $\acute{g_2} m_i \delta \acute{x_i}$, where $m_i$ and $\delta \acute{x_i}$ respectively break time reversal and inversion symmetry. We remedy this situation by including in our ansatz the additional assumption that $\acute{g_2}$ is proportional to the antiferromagnetic order parameter $L$ which breaks both inversion and time reversal symmetry. Accordingly, since both $\alpha$ and $\acute{g_2}$ both have the same symmetry properties of $L$, the interaction contribution,
\be
F_{int}= -\mu_0\alpha PM = -\acute{g_1} \acute{g_2}PM /2\acute{k}  ,
\label{strainFE_v3}
\ee
includes a strain contribution which by this simple qualitative model is symmetry preserving. 

A hidden antiferromagnetic order parameter for magnetoelectric BMO is postulated by Solovyev and Pchelkina\cite{hiddenAFM} using microscopic theory in which a relativistic spin-orbit interaction gives rise to canted spin ferromagnetism. In a separate and more recent theoretical treatment using hybrid density functional theory, a highly-strained antiferromagnetic supertetragonal phase is predicted to have a high polarization\cite{supertetragonal}. These two theories of improper multiferroicity in BMO\cite{hiddenAFM, supertetragonal} thus allow for the coexistence and interplay of ferroelectricity and ferromagnetism in a system which is otherwise centrosymmetric (C2/c) and incompatible with ferroelectricity. 

This scenario is made more plausible with the realization that ferromagnetism in BiMnO$_{3}$ is due to the orbital ordering of the spin carrying Mn atoms that is sensitive to atomic spacings\cite{BMO_OO}. In addition the magnetization, with a maximum of $\sim 1 \mu_B$/Mn\cite{Jeen} in our films, is significantly reduced from the bulk value of 3.6\thinspace $\mu_B$/Mn, suggesting that the magnetization is inhomogeneous and likely contains a strain-induced antiferromagnetic component. The atomic spacings are modified by the high strain island edges\cite{Chen}, misaligning the magnetic moments, resulting in a spatially varying magnetization across the islands which in turn produces internal electric fields\cite{ME3} that are capable of aligning electric dipoles. There may also be an additional contribution to the measured magnetoelectric coefficient $\tilde{\alpha}(T)$ (Fig.~\ref{figure4}) from the overall reduction in volume of the ferroelectric regions on the application of a magnetic field, thus increasing the magnitude of the negative magnetoelectric coupling.

Temperature dependent high-pressure neutron diffraction studies have also reported a pressure induced monoclinic-to-monoclinic structural transition which leads to a transformation from a ferromagnetic to antiferromagnetic phase\cite{Kozlenko}. It is not difficult to imagine that this competition and sensitivity of magnetic order parameters to slightly different structures, albeit at high pressure, is present in strain-disordered thin films at atmospheric pressure.    

\subsection{Bilinear interaction and enhanced magnetoelectric coupling}

The omission of quartic terms proportional to $P^4$ and $M^4$ and a biquadratic coupling term proportional to $(PM)^2$ in Eq.~\ref{Helmholtz3} considerably simplifies our analysis. 
Applying the thermodynamic expressions $E = (\partial F/\partial P)_{T,M}$ and $H = \mu_0^{-1}(\partial F/\partial M)_{T,P} $ to Eq.~\ref{Helmholtz3}, a matrix inversion leads to expressions for $P(E,H)$and $M(E,H)$ that are linear in the applied fields $E$ and $H$. The result gives the expressions 
\bea
\delta P &=& P - P_R = \epsilon_0 \tilde{\chi}_P(T)E + \tilde{\alpha}(T) H \cr
\cr
\mu_0 \delta M &=&\mu_0 (M - M_R) = \tilde{\alpha}E + \mu_0 \tilde{\chi}_M(T)H
\label{observables}
\eea
where $P_R$ and $M_R$ are the remnant moments when the applied fields $E$ and $H$ are zero. The renormalized susceptibilities $\tilde{\chi}_P(T) = \chi_P/(1 - \alpha^2 c^{-2}\chi_P(T)\chi_M(T))$ and $\tilde{\chi}_M(T) = \chi_M(T)/(1 - \alpha^2 c^{-2}\chi_P(T)\chi_M(T))$ together with the magnetoelectric coeficient 
\bea
\tilde{\alpha}(T) &=& \alpha c^{-2}\chi_P(T)\chi_M(T)/(1 - \alpha^2 c^{-2}\chi_P(T)\chi_M(T)) \cr
\cr
&=& \alpha/(c^2\chi_P^{-1}(T)\chi_M^{-1}(T) - \alpha^2)
\label{alphatwiddle}
\eea
are the experimental observables. We note that $P$ and $\tilde{\alpha}(T) H$ in the first line of Eq.~\ref{observables} both break inversion symmetry since the time reversal asymmetry in $\tilde{\alpha}(T)$ and $H$ appears twice and therefore cancels. A similar argument applies to the $\tilde{\alpha}(T) E$ term in the second line of Eq.~\ref{observables} which breaks time reversal symmetry with the preservation of inversion symmetry.

\begin{figure}
\includegraphics[angle=0,width=0.5\textwidth]{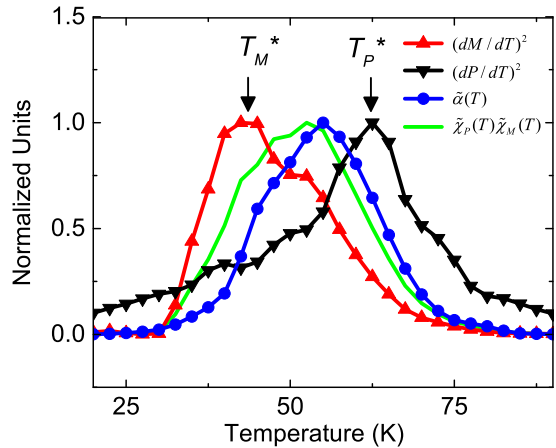}
\caption{The temperature derivatives of the magnetization (red) and ferroelectric polarization (black) are calculated from Fig. 2 and provide a measure of the magnetic and electric susceptibilities (see text). The ME coupling coefficient, $\tilde{\alpha} (T)$, is shown in blue, and is independently calculated from the slopes of $P_{R}$ vs. $B$ curves at each temperature (see Fig. 4. The green line representing the normalized product $\tilde{\chi}_M (T) \tilde{\chi}_P (T)$ (determined by the independently measured temperature derivatives of Eqs.~\ref{susceptibilities}) correlates well with the measured values of $\tilde{\alpha} (T)$ in accordance with Eq.~\ref{peakoverlays}.}
\label{figure5}
\end{figure}

Parenthetically, we note from Eq.~\ref{alphatwiddle} above that $\alpha$ and $\tilde{\alpha}$ have the same positive sign determined by the relation $\alpha = a_1 L/\mu_0$ where $a_1$ in Eq.~\ref{Helmholtz2} is defined as positive. If $a_1$ in Eq.~\ref{Helmholtz2} is replaced by $-a_1$, the free energy minimum occurs at $\theta_0 = \pi/2$ rather than $\theta_0 = 3\pi/2$ and Eq.~\ref{Helmholtz3} remains unchanged.
The experimentally observed decrease in $P_R$ with applied field $H$, i.e., $\partial P_R/\partial H = -\tilde{\alpha}$ is thus consistent with the signs in Eq.~\ref{observables}.

The materials-specific coupling constant $\alpha$ has dimensions corresponding to velocity and gives rise to a linear magnetoelectric effect\cite{Revival}. If there is no ME coupling, then $\alpha$ in Eq.~\ref{Helmholtz3} is equal to zero as is $\tilde{\alpha}(T)$. In this case there are two separate and independent phase transitions with the higher temperature ferroelectric transition occurring at $T = T_P$ where $\chi_P^{-1} = (T/T_P - 1)/C_P$ goes to zero and the lower temperature ferromagnetic transition occurring at $T = T_M$ where $\chi_M^{-1} = (T/T_M - 1)/C_M$ goes to zero. The Curie temperature prefactors, $C_P$ and $C_M$, are taken to be dimensionless. 

In the presence of finite coupling, the essential physics is markedly different as captured in the behavior of the determinant $c^2\chi_P^{-1}\chi_M^{-1} - \alpha^2= \alpha/\tilde{\alpha}(T)$ associated with the denominator of Eq.~\ref{alphatwiddle}, which with substitution of the expressions in the preceding paragraph for the temperature-dependent susceptibilities becomes
\bea
(T-T_P)(T-T_M) - \alpha^2 C_PC_MT_PT_M/c^2 = \cr
\cr
(T-T_P^{\ast})(T-T_M^{\ast}) = \alpha C_PC_MT_PT_M/c^2 \tilde{\alpha}(T) .
\label{denominator}
\eea
By factoring the left most term of the above equation into the product  $(T-T_P^{\ast})(T-T_M^{\ast})$ the quadratic equation in $T$ can be solved for the two roots
$T_P^{\ast}$ and $T_M^{\ast}$ as,
\bea
&&T_P^{\ast} , T_M^{\ast} = [T_p + T_M \pm ( T_P + T_M ) \cr
\cr
&&(1 + 4\alpha^2C_PC_MT_PT_M/c^2(T_P - T_M)^2)^{1/2}]/2
\label{quadratic}
\eea

In the absence of magnetoelectric coupling, $\alpha = 0$ and the renormalized temperatures $T_P^{\ast}$ and $T_M^{\ast}$ are unchanged from the original separate and independent phase transitions respectively at $T=T_P$ and $T=T_M$. If $T_P > T_M$ in the uncoupled system, as is the case in our experiment, then as coupling increases Eq.~\ref{quadratic} shows that $T_P^{\ast} - T_M^{\ast}$ also increases with $T_P^{\ast} > T_P$ and $T_M^{\ast} < T_M$. A similar repulsion (increase in separation) occurs if $T_P < T_M$. It is important to recognize that the highest of the two temperatures, $T_P^{\ast}$, marks a susceptibility transition where both interacting electric and magnetic dipoles exist. It does not make sense to identify the lower transition at $T = T_M^{\ast}$ as a phase transition temperature; any remnant feature at this temperature, such as a susceptibility peak, is a byproduct of interactions which have their onset at the higher temperature $T_P^{\ast}$.  

Interpretation of the meaning of the renormalized temperatures $T_P^{\ast}$ and $T_M^{\ast}$ is further clarified by solving the rightmost equality of
Eq.~\ref{denominator} for the observable $\tilde{\alpha}(T)$, i.e., 
\be  
\tilde{\alpha}(T) = \alpha c^{-2} C_PC_MT_PT_M/(T-T_P^{\ast})(T-T_M^{\ast})
\label{alphatwiddle1}
\ee
and comparing the denominator of this result with the denominator of Eq.~\ref{alphatwiddle}. The magnetoelectric coupling constant $\tilde{\alpha} (T)$ reaches a maximum at temperatures where both denominators are small, i.e., when  $(c^2\chi_P^{-1}\chi_M^{-1}-\alpha ^2) \approx 0)$ (Eq.~\ref{alphatwiddle}) or equivalently when
$(T-T_P^{\ast})(T-T_M^{\ast}) \approx 0$ (Eq.~\ref{alphatwiddle1}). By writing the proportionalities, $\tilde{\chi}_P(T) \propto \vert T-T_P^{\ast} \vert^{-1}$ and $\tilde{\chi}_M(T) \propto \vert T-T_M^{\ast}\vert^{-1}$\thinspace suggested by the discussion following Eqs.~\ref{observables}, Eq.~\ref{alphatwiddle1} becomes a proportionality, 
\be
\tilde{\alpha}(T) \propto \tilde{\chi}_P(T) \tilde{\chi}_M(T) 
\propto  \vert T-T_P^{\ast}\vert ^{-1}\vert T-T_M^{\ast}\vert ^{-1} ~~,
\label{peakoverlays}
\ee
showing that the magnetoelectric coefficient $\tilde{\alpha}(T)$ has the same temperature dependence as the product $\tilde{\chi}_P(T) \tilde{\chi}_M(T)$ of susceptibilities. 

In the final steps leading to Eq.~\ref{peakoverlays} we have replaced the parentheses around the temperature differences by absolute values, since the susceptibilities as calculated by Eq.~\ref{proportionalities} are proportional to well-behaved continuous derivatives of smooth transitions shown in Fig.~\ref{figure1}. We have also assumed that $L \propto |\vec{M}|$ so that the normalized temperature dependence of $LM$ is close to that of $M$. Said in another way, the magnetization transition shown in Fig.~\ref{figure1}~includes the temperature dependence of the hidden order parameter $L$.    

\begin{figure}
\includegraphics[angle=0,width=0.5\textwidth]{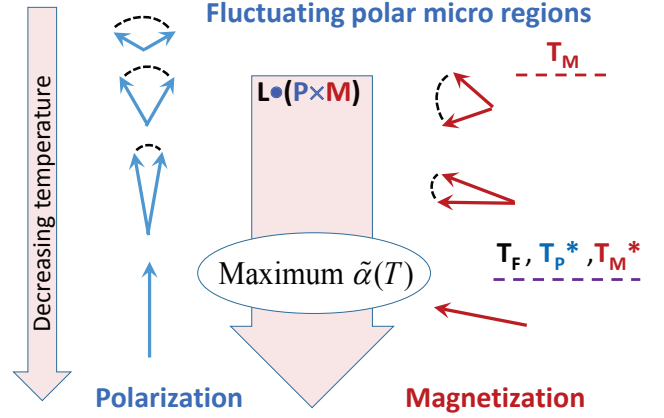}
\caption{Schematic showing evolution of polarization P (blue arrows) and magnetization M (red arrows) as a function of decreasing temperature $T$ (vertical axis). For $T>T_M$, where $T_M = 105$~K is the unrenormalized bulk transition temperature, the rapidly fluctuating electric dipoles of the polar microregions of the relaxor ferroelectric dominate. As $T$ decreases the dipoles increase in size and relaxation times (fluctuation rates) increase (decrease). At $T = T_M$ magnetic moments appear, which because of the trilinear coupling, $\vec{L} \cdot (\vec{P} \times \vec{M})$, fluctuate at rates comparable to the relaxor rates. As $T$ is lowered still further towards the relaxor freezing temperature $T_F \approx 70$~K, the electric and magnetic dipoles simultaneously maintain their coupled orientations on laboratory time scales at comparable transition temperatures, $T_F \approx T_P^{\ast} \approx T_M^{\ast}$, where the magnetoelectric coupling coefficient, $\tilde{\alpha}(T)$, is maximum. As $T$ decreases further, both $P$ and $M$ continue to increase as $\tilde{\alpha}(T)$ begins to decrease (see Figs.~\ref{figure1} and~\ref{figure5}). As discussed in the text the free energy is minimized when the alignment between $\vec{P}$ and $\vec{M}$ is $3\pi/2$; the misalignment shown in the figure schematically illustrates the canting away from $3\pi/2$ introduced by the presence of an antiferromagnetic component, $\vec{L}$.}  
\label{figure6}
\end{figure}

Agreement with the overlapping temperature dependences predicted by Eq.~\ref{peakoverlays} is shown in Fig.~\ref{figure5} where the separately measured susceptibilities, $\tilde{\chi}_P(T)$ (solid black inverted triangles) and $\tilde{\chi}_M(T)$ (solid red triangles), and magnetoelectric coefficient $\tilde{\alpha}(T)$ (solid blue circles) are plotted with respective maxima normalized to unity. The susceptibility peaks, which occur at the temperatures $T_P^{\ast}$ and $T_M^{\ast}$ ($T_P^{\ast} > T_M^{\ast}$) are marked with vertical arrows and are seen to symmetrically straddle the peak in $\tilde{\alpha}(T)$. The green line representing the computed normalized product $\tilde{\chi}_M (T) \tilde{\chi}_P (T)$ correlates well with the measured values of $\tilde{\alpha} (T)$. The data show very clearly, in accordance with the result of Eq.~\ref{peakoverlays} how the overlapping portions of the proximate susceptibility transitions amplify magnetoelectric coupling. If the biquadratic interaction term [$\propto (PM)^2$] alone is used in the free energy, the ME effect can also be linear in $H$ but only exists below the lowest $T_c$ \cite{boyd, Agyeit}, in contrast to the experimental results in Fig.~\ref{figure5} where $\tilde{\alpha}$ is finite at temperatures well above the susceptibility peak at $T = T^{\ast}_M$.

\section{Summary and Conclusions}

In summary, our study of epitaxial thin films of BiMnO$_3$ has uncovered a robust ferroelectric remnant polarization ($23 \mu$C/cm$^2$ at 5~K) with a pronounced magnetoelectric coefficient (-1.25 ns/m at 55~K), neither of which are present in bulk crystalline BiMnO$_3$. 
The polarization and magnetization transitions are in close proximity with the temperature dependence of the product of susceptibilities closely following the temperature dependence of the magnetoelectric coefficient. This proximity of transitions gives a strongly enhanced linear magnetoelectric coupling. Our use of a symmetry preserving trilinear coupling term of the form $\vec{L} \cdot (\vec{P} \times \vec{M})$ enables a thermodynamic interpretation in which, after angle averaging, a bilinear term proportional to PM in the free energy provides an excellent qualitative description of the data in which there is only one phase transition marked by simultaneous maxima in the cross susceptibility and magnetoelectric coefficient (Eq.~\ref{alphatwiddle1}). Although experimentally we measure the suppression of a remnant polarization, our analysis does not require the inclusion of quartic or biquadratic terms in the free energy to determine the remnant moments and how they are affected by the application of external fields. Rather, the inclusion of a trilinear interaction in the free energy suffices to give a good account of the observed large linear magnetoelectric coupling. 

Strain disorder clearly plays an important role in understanding the multiferroicity and magnetoelectric coupling in BiMnO$_3$.
The ferroelectricity is of a relaxor origin with relaxations due to thermally activated reorientation of disorder-induced 
polar micro regions observable (see Fig.~\ref{relaxorloops}) at temperatures greater than $\approx 90$~K.
At temperatures below the relaxor freezing temperature of $\approx$ 70K, the  remnant polarizations are stable and exhibit a pronounced sensitivity to the application of externally applied strains\cite{Mickel1}.
We find that the distortions associated with the polar micro regions are described by a local strain variable, which in our simple model couples linearly to both $P$ and $M$, 
giving rise to the requisite bilinear interaction term proportional to $PM$ in the free energy. These very same distortions are responsible for the perturbations in the orbital ordering of the spin carying Mn atoms that are responsible for the antiferromagnetic component $\vec{L}$. The Dzyaloshinskii-Moriya interaction, represented in our analysis by the symmetry conserving interaction term $\vec{L} \cdot (\vec{P} \times \vec{M})$, favors non-uniform magnetic structures that are an essential ingredient to the observed magnetoelectric coupling 

Our analysis leaves two open issues needing further research. The first of these is the absence of convincing magnetic evidence (other than a suppressed saturated moment) for a hidden antiferromagnetic order parameter as proposed for the interaction term in Eq.~\ref{Helmholtz2} and discussed in section \ref{subsec:a hidden antiferromagnetic order parameter}. The review by Belik\cite{BMOreview} cites a number of papers reporting ferroelectricity in high resistivity BMO films\cite{DosSantos,Sharan,Son,Gajek,STObuf,Jeen,DeLuca} occurring concomitantly with a reduced saturated moment compared to the bulk value of 3.6$\mu_B$/Mn. Interestingly, a recent publication on high quality pulse laser deposited BMO films with good stoichiometry reports high resistivity together with a ferromagnetic $T_c = 105$~K and a saturated moment of 3.6$\mu_B$/Mn as seen in bulk, but no evidence of ferroelectricity\cite{Jung}. This evidence, though circumstantial, supports our conclusion that multiferroic behavior of BMO films requires the presence of an antiferromagnetic component which acts to reduce the saturated moment. On another experimental front, neutron scattering measurements could resolve an antiferromagnetic component within a ferromagnetic background, but such measurements are difficult because the ferroelectricity and large magnetoelectric effect in BMO only appears in thin strained films or at interfaces, not in bulk. 

A second open issue concerns the question of whether the proximity of the transition temperatures is accidental. We have shown that in the presence of symmetry preserving trilinear coupling, there is only one phase transition and both $P_R$ and $M$ appear at that temperature. Clearly a time stable magnetoelectric effect cannot be present until the temperature is low enough (i.e., $ T < T_F = 70$~K to have stable dipole moments. At first sight there is no such restriction on the magnetic transition at higher temperatures which in bulk appears at a Curie transition $T_c = 105$~K. One possibility, depicted schematically in Fig.~\ref{figure6}, is that the coupling of the rapidly fluctuating dipole moments to magnetic moments in this higher temperature region ($T > T_F$) is sufficiently strong to prevent time-stable magnetization. Accordingly, in this very plausible scenario the transitions in $P_R$ and $M$ appear simultaneously near $T = T_F$, where the relaxation time of the ferroelectric diverges and both of the coupled moments are stable, as is observed experimentally.

In conclusion, our study of a surprisingly large linear magnetoelectric effect in BiMnO$_3$ opens a new perspective on the design of novel multiferroics that might eventually have practical application. The combination of strain, magnetic ordering and relaxor ferroelectricity conspire to bring the transition temperatures of the respective ferroic order parameters into close proximity with each other, thereby significantly enhancing the magnetoelectric coupling. In addition, our theoretical treatment utilizing a Dzyaloshinskii-Moriya type trilinear interaction avoids the use of higher order (and hence smaller) biquadratic interaction terms and gives a good account of the enhanced ME coefficient.     

\section*{Acknowledgements}
The authors thank Sanal Buvaev for assistance with sample preparation and measurements and D. Maslov and M. Mostovoy for useful discussions on points of theory. This work was supported by the U. S. National Science Foundation under Grant Nos. DMR-1305783 (AFH) and DMR-0804452 (AB).




\begin{thebibliography}{30}
\bibdata

\bibitem{Revival} M. Fiebig, \textit{Revival of the magnetoelectric effect}, J. Phys. D: Appl. Phys. \textbf{38}, R123 (2005).

\bibitem{ME1} R. Ramesh, and N. A. Spaldin, \textit{Multiferroics: progress and prospects in thin films}, Nature Mater. \textbf{6}, 21 (2007).

\bibitem{ME3} S.W. Cheong, and M. Mostovoy, \textit{Multiferroics: a magnetic twist for ferroelectricity}, Nature Mater. \textbf{6}, 13 (2007).

\bibitem{ME60s} W. F. Brown, R.M. Hornreich, and S. Shtrikman, \textit{Upper bound on the magnetoelectric susceptibility}, Phys. Rev. \textbf{168}, 574 (1968).

\bibitem{ME2} W. Erenstein, N. D. Mathur, and J.F. Scott, \textit{Multiferroic and magnetoelectric materials}, Nature \textbf{442}, 17 (2006).

\bibitem{dzya} I. Dzyaloskinskii, \textit{Magnetoelectric to multiferroic phase transitions}, Eur. Phys. Lett. \textbf{96}, 17001 (2011).

\bibitem{d0} N.A. Hill,  \textit{Why are there so few magnetic ferroelectrics?} J. Phys. Chem. B \textbf{104}, 669709 (2000).

\bibitem{boyd} G. R. Boyd, P. Kumar, and S. R. Phillpot, \textit{Multiferroic Materials and their Properties}, arXiv 1101.5403; P. Kumar, Integrated Ferroelectrics: An International Journal \textbf{131}, 25 (2011).

\bibitem{HillRabe} N. A. Hill, and K. M. Rabe, \textit{First-principles investigation of ferromagnetism and ferroelectricity in bismuth manganite}, Phys. Rev. B \textbf{59}, 8759 (1999).

\bibitem{BMO_OO} A. M. dos Santos, A. K. Cheetham, T. Atou, Y. Syono, Y. Yamaguchi, K. Ohoyama, H. Chiba, and C. N. R. Rao, \textit{Orbital ordering as the determinant for ferromagnetism in biferroic BiMnO$_{3}$}, Phys. Rev. B \textbf{66}, 064425 (2002).

\bibitem{Vis6s} R. Seshadri, and N. A. Hill, \textit{Visualizing the role of Bi 6s lone pairs in the off-center distortion in ferromagnetic BiMnO$_{3}$}, Chem. Mater. \textbf{13}, 2892899 (2001).

\bibitem{Baettig} P. Baettig, R. Seshadri, and N. A. Spaldin, \textit{Anti-polarity in ideal BiMnO$_3$}, J. Am. Chem. Soc. \textbf{129}, 9854 (2007).

\bibitem{Belik} A. A. Belik, et al., \textit{Origin of the monoclinic-to-monoclinic phase transition and evidence for the centrosymmetric crystal structure of BiMnO$_3$}, J. Am. Chem. Soc. \textbf{129}, 971-977 (2007).

\bibitem{Scott2} W. Eerenstein, F. D. Morrison, F. Sher F,  J. L. Prieto, J. P. Attfield, J. F. Scott, and N. D. Mathur, \textit{Experimental difficulties and artifacts in multiferroic and magnetoelectric thin films of BiFeO$_3$, Bi$_{0.6}$Tb$_{0.3}$La$_{0.1}$FeO$_{3}$ and BiMnO$_{3}$}, Phil. Mag. Lett. \textbf{87}, 249 (2007).

\bibitem{Sharan} A. Sharan, J. Lettieri, Y. Jia, W. Tian, X. Pan, D. G. Schlom, and V. Gopalan, \textit{Bismuth manganite: A multiferroic with a large nonlinear optical response}, Phys. Rev. B \textbf{69}, 214109 (2004).

\bibitem{DosSantos} A. Moreira dos Santos, S. Parashar, A.R. Raju, Y.S. Zhao, A.K. Cheetham, C.N.R. Rao, \textit{Evidence for the likely occurrence of magnetoferroelectricity in the simple perovskite, BiMnO$_3$}, Solid State Commun. \textbf{122}, 49 (2002).


\bibitem{Son} J. Y. Son, Bog G. Kim, C. H. Kim and J. H. Cho, \textit{Writing polarization bits on the multiferroic BiMnO$_3$ thin film using Kelvin probe force microscope}, Appl. Phys. Lett. \textbf{84}, 4971 (2004).

\bibitem{Gajek} M. Gajek, M. Bibes, F. Wyczisk, M. Varela, J. Fontcuberta, and A. Barthélémy, \textit{Growth and magnetic properties of multiferroic La$_x$Bi$_{1-x}$MnO$_3$ thin films}, Phys. Rev. B \textbf{75}, 174417 (2007).

\bibitem{STObuf} J. Y. Song, and Y-H. Shin, \textit{Multiferroic BiMnO$_3$ thin films with double SrTiO$_3$ buffer layers}, Appl. Phys. Lett. \textbf{93}, 062902 (2008).

\bibitem{Jeen} H. Jeen, G. Singh-Bhalla, P. R. Mickel, K. Voight, C. Morien, S. Tongay, A. F. Hebard, and A. Biswas, \textit{Growth and characterization of multiferroic BiMnO$_3$ thin films}, J. Appl. Phys. \textbf{109}, 074104 (2011).

\bibitem{DeLuca} G. M. De Luca, D. Preziosi, F. Chiarella, R. Di Capua, S . Gariglio, S. Lettieri and M. Salluzso, \textit{Ferromagnetism and ferroelectricity in epitaxial BiMnO$_3$ ultra-thin films}, Appl. Phys. Lett. \textbf{103}, 062902 (2013).

\bibitem{BMOreview} A. A. Belik, \textit{Polar and nonpolar phases of BiMnO$_3$: A review}, J. Solid State Chem. \textbf{195}, 32 (2012).

\bibitem{Fennie} C. J. Fennie, \textit{Ferroelectrically induced weak ferromagnetism by design}, Phys. Rev. Lett. \textbf{100}, 167203 (2008); Also see, C. Ederer and C. J. Fennie, \textit{Electric-field switchable magnetization via the Dzyaloshinskii-Moriya interaction: FeTiO$_3$ versus BiFeO$_3$}, J. Phys. Cond. Mat. \textbf{20}, 434219 (2008).

\bibitem{Dzy} I. Dzyaloshinskii, \textit{A thermodynamic theory of “weak”
ferromagnetism of antiferromagnets}, J. Phys. Chem. Solids \textbf{4}, 241 (1958).

\bibitem{Moriya} T. Moriya, \textit{Anisotropic Superexchange Interaction and Weak Ferromagnetism}, Phys. Rev. \textbf{120}, 91 (1960).

\bibitem{hiddenAFM} I. V. Solovyev and Z. V. Pchelkina, \textit{Magnetic-field control of the electric polarization in BiMnO$_3$}, Phys. Rev. B \textbf{82}, 094425 (2010).

\bibitem{supertetragonal} O. Di\'{e}gueez and J. \'{I}\~{n}iguez, \textit{Epitaxial phases of BiMnO$_3$ from first principles}, arXiv: 1503.08293 (2015).

\bibitem{Chen} Y. Chen, and J. Washburn, \textit{Structural Transition in Large-Lattice-Mismatch Heteroepitaxy}, Phys. Rev. Lett. \textbf{77}, 4046 (1996).

\bibitem{Biswas} A. Biswas, M. Rajeswari, R. C. Srivastava, Y. H. Li, T. Venkatesan, R. L. Greene, and A. J. Millis, \textit{Two-phase behavior in strained thin films of hole-doped manganites}, Phys. Rev. B \textbf{61}, 9665 (2000).

\bibitem{Chiba} H. Chiba, T. Atou, and Y. Syono, \textit{Magnetic and Electrical properties of Bi$_{1-x}$Sr$_x$MnO$_3$: Hole-doping effect on ferromagnetic perovskite BiMnO$_3$}, J. Solid State Chem. \textbf{132}, 139 (1997).

\bibitem{Mickel1} P. R. Mickel, Hyoungjeen Jeen, A. Biswas and A. F. Hebard, \textit{Orientational Strain Modification of Ferroelectric Polarization in Multiferroic BiMnO$_3$}, J. Appl. Phys. \textbf{105}, 262904 (2014).

\bibitem{Vogel} R. Pirc, and R. Blinc, \textit{Vogel-Fulcher freezing in relaxor ferroelectrics}, Phys. Rev. B \textbf{76}, 020101 (2007).
 
\bibitem{Ang} Ang, C., Yu, Z., Lunkenheimer, P., Hemberger, J., and Loidl, A., \textit{Dielectric relaxation modes in bismuth-doped SrTiO$_3$: The relaxor behavior}, Phys. Rev. B \textbf{59}, 6670 (1999).

\bibitem{Tsurumi} T. Tsurumi, K. Soejima, T. Kamiya, and M. Daimon, \textit{Mechanism of Diffuse Phase-Transition in Relaxor Ferroelectrics}, Jpn. J. Appl. Phys. \textbf{33}, 1959 (1994).

\bibitem{TbPO4} G. T. Rado, J. M. Ferrari, and W. G. Maisch, \textit{Magnetoelectric susceptibility and magnetic symmetry of magnetoelectrically annealed TbPO$_{4}$}, Phys. Rev. B\textbf{29}, 4041 (1984).

\bibitem{Spirals} Y. Tokura, and S. Seki, \textit{Multiferroics with spiral spin orders}, Adv. Mater. \textbf{22}, 1554565 (2010).

\bibitem{Scott} A. Kumar, G. L. Sharma, R. S.Katiyar, R. Pirc, R. Blinc,
and J F Scott, \textit{Magnetic control of large
room-temperature polarization}, J. Phys.: Condens. Matter \textbf{21}, 382204(2009).

\bibitem{Mikaelyan} M. A. Mika\'{e}lyan, \textit{Methodological aspects of the thermodynamics of dielectrics}, Physics-Uspekhi \textbf{41}, 1219 (1998).

\bibitem{Rushbrooke} G. S. Rushbrooke, \textit{On the thermodynamics of the critical region for the Ising problem}, J. Chem. Phys. \textbf{39}, 842 (1963).

\bibitem{Binney} J. J. Binney, N. J. Dowrick, A. J. Fisher, and M. E. J. Newman, \textit{The Theory of Critical Phenomena: An Introduction to the Renormalization Group}, Oxford Science Publications (Clarendon Press) Oxford (1992).

\bibitem{Howczak} O. Howczak and J. Spalek, \textit{Ferroelectric-ferromagnetic correlations in BiMnO$_3$ perovskite within Landau theory: comparison with experiment}, Eur. Phys. J. B \textbf{78}, 417 (2010).

\bibitem{Kozlenko} D. P. Kozlenko, A. A. Belik, S. E. Kichanov, I. Mirebeau, D. V. Sheptyakov, Th. Strassle, O. L. Makarova, A. V. Belushkin, B. N. Savenko, and E. Takayama-Muromachi, \textit{Competition between ferromagnetic and antiferromagnetic ground states in multiferroic BiMnO$_3$ at high Pressures}, Phys. Rev. B \textbf{82}, 014401 (2010).
 
\bibitem{Agyeit} A. K. Agyeit and J. L. Birman, \textit{On the linear magnetoelectric effect}, J. Phys.: Condens. Matter \textbf{2}, 3007 (1990).

\bibitem{Jung} B. W. Lee, P. S. Yoo, V. B. Nam, K. R. N. Toreh and C. U. Jung, \textit{Magnetic and electric properties of stoichiometric BiMnO$_3$ films}, Nanoscale Res. Lett. \textbf{10}, 47 (2015).











\end{thebibliography}
\end{document}